\documentclass[floatfix,aps,twocolumn,preprintnumbers,amsmath,amssymb,nofootinbib,superscriptaddress]{revtex4}

\usepackage{graphicx,color}
\usepackage{amsmath,amssymb,bm}
\usepackage{natbib} 

\def\1{\'{\i}}

\def\XXint#1#2#3{{\setbox0=\hbox{$#1{#2#3}{\int}$}
     \vcenter{\hbox{$#2#3$}}\kern-.5\wd0}}

\newcommand{\nk}{{\bf k}}

\newcommand{\nq}{{\bf q}}

\def\XXint#1#2#3{{\setbox0=\hbox{$#1{#2#3}{\int}$}
     \vcenter{\hbox{$#2#3$}}\kern-.5\wd0}}

\def\1{\'{\i}}

\begin{document}

%\begin{frontmatter}
\title{Axial-vector dominance predictions in quasielastic
neutrino-nucleus scattering}

\author{J.E. Amaro}\email{amaro@ugr.es} \affiliation{Departamento de
  F\'{\i}sica At\'omica, Molecular y Nuclear \\ and Instituto Carlos I
  de F{\'\i}sica Te\'orica y Computacional \\ Universidad de Granada,
  E-18071 Granada, Spain.}
  
\author{E. Ruiz
  Arriola}\email{earriola@ugr.es} \affiliation{Departamento de
  F\'{\i}sica At\'omica, Molecular y Nuclear \\ and Instituto Carlos I
  de F{\'\i}sica Te\'orica y Computacional \\ Universidad de Granada,
  E-18071 Granada, Spain.}

\begin{abstract}

The axial form factor plays a crucial role in quasielastic
neutrino-nucleus scattering, but the error of the theoretical cross
section due to uncertainties of $G_A$ remains to be established.
Reversely, the extraction of $G_A$ from the neutrino nucleus cross
section suffers from large systematic errors due to nuclear model
dependencies, while the use of single parameter dipole fits
underestimates the errors and prevents an identification of the
relevant kinematics for this determination. We propose to use a
generalized axial-vector-meson-dominance (AVMD) in conjunction with
large-$N_c$ and high energy QCD constrains to model the nucleon axial
form factor, as well as the half width rule as an a priori uncertainty
estimate.  The minimal hadronic ansatz comprises the sum of two
monopoles corresponding to the lightest axial-vector mesons being
coupled to the axial current.  The parameters of the resulting axial
form factor are the masses and widths of the two axial mesons as
obtained from the averaged PDG values.  By applying the half width
rule in a Monte Carlo simulation, a distribution of theoretical
predictions can then be generated for the neutrino-nucleus
quasielastic cross section.  We test the model by applying it to the
$(\nu_\mu,\mu)$ quasielastic cross section from $^{12}$C for the
kinematics of the MiniBooNE experiment.  The resulting predictions
have no free parameters.  We find that the relativistic Fermi gas
model globally reproduces the experimental data, giving $\chi^2/ \#
bins = 0.81$. A $Q^2$-dependent error analysis of the neutrino data
shows that the uncertainties in the axial form factor $G_A(Q^2)$ are
comparable to the ones induced by the a priori half width rule. We
identify the most sensitive region to be in the range $0.2 \lesssim
Q^2 \lesssim 0.6 \,{\rm GeV}^2$.
\end{abstract}

%\pacs{24.10.Jv,25.30.Fj,25.30.Pt,21.30.Fe} 

%\keywords{Neutrino reactions, quasielastic electron scattering, axial form factor, axial mesons, relativistic Fermi gas, Monte Carlo simulation}

%\end{frontmatter}

\date{\today}



\maketitle

\section{Introduction}

Since the first measurement of the muon neutrino charged current
quasielastic double differential cross section
\cite{Agu08,Agu10,Agu13} many attempts have been made to characterize
an effective axial-vector form factor of the nucleon
\cite{Gal11,For12,Mor12,Alv14}.  This is often made in terms of a
dipole axial mass $M_A$, assuming a dipole form \cite{Nie12,But14},
for $Q^2 >0$
\begin{eqnarray}
G_A^{\rm  dipole}(Q^2) &=& \frac{g_A}{(1+Q^2/M^2_A)^2} \, .
\label{eq:dipole}
\end{eqnarray}
The world average value of the nucleon dipole axial mass is $M_A\sim
1$ GeV \cite{Liesenfeld:1999mv}
(see e.g. \cite{Ber02} for a review and references therein)
which is obtained as a weighted sum of different oncoming dipole
fits to independent experiments. However, this does not mean that the
full spread of axial form factors can be described by a single dipole
mass with a given uncertainty in a statistically significant
way. Actually, there is a great variety of often largely incompatible
experimental data from different processes which for illustration can
be seen at Fig.~\ref{fa1}. The correct discrimination and selection of
these mutually compatible data is a complicated problem in data
analysis which requires proper weighting of experimental ranges and
falsifiable reliable theoretical input and not just parameterizations
which awaits resolution and will not be addressed here. Nonetheless,
given the present rather confusing state of affairs and the lack of
further qualified information on what data on the axial form factor
should one objectively prefer we will face the problem from a
different and somewhat unconventional perspective where the
traditional fitting strategy is sidestepped by the use of a
theoretically based axial form factor with an inherent error band.

The MiniBooNE cross section data values are too large compared to the
theoretical models of quasielastic neutrino scattering in the impulse
approximation, unless a significant larger value of $M_A \sim 1.35$
GeV is employed in the nuclear axial current. Microscopic explanations
of the large value of $M_A$ have been proposed based on ingredients
involving nucleon spectral functions and multinucleon emission
induced by short range correlations and meson exchange currents
\cite{Mar10,Ama11,Nie12}. Recently studies with a monopole
parameterization have been performed in \cite{Meg13} as well as nucleon
mean field effective mass analyses of blurred electron scattering
data~\cite{Amaro:2015zja}.  In the absence of reliable theoretical
uncertainty estimates, the disparate values obtained upon
consideration of different nuclear effects can so far be regarded as a
genuine source of systematic errors. Those turn out to be much larger
than the alleged statistical uncertainties which, if taken literally,
would lead to the most precise determination of the axial form factor
to date in the range $Q^2 \lesssim 2 \,{\rm GeV}^2$. A careful
statistical analysis has been undertaken more
recently~\cite{Wilkinson:2015gea} and some tension among different
data in different models has been reported.

The popular dipole form factor parameterization enjoys the pQCD
result~\cite{Carlson:1985zu} asymptotically, $G_A \sim 1/Q^4$, but
despite the phenomenological success for separate and independent
experiments, it finds no further theoretical support at finite $Q^2$,
nor does it describe {\it all} experiments globally with an acceptable
$\chi^2$ value. Moreover, a one parameter fit such as the dipole form
introduces an artificial bias linking high and low energies
unnaturally and tightly; it is unclear if the statistical fluctuations
inherited from the uncertainties in experimental neutrino-nucleus
scattering data are faithfully represented by the corresponding
fluctuations in the dipole mass. This is a well known issue in the
statistical analysis of data since the goodness of fit and the
parameter confidence level is based on estimating the probability that
the proposed parameterization be the correct one, and this implies a
mapping between data fluctuations and the fitting parameter
fluctuations.  To overcome this limitation a model independent
analysis of axial form factor using dispersion relations under
definite convergence assumptions and based on neutrino scattering was
performed~\cite{Bhattacharya:2011ah}, with the expected finding that
errors inferred from a dipole ansatz analysis may be underestimated. A
duality based parameterization has been proposed searching for
significant deviations to the widely used dipole
form~\cite{Bodek:2007ym}. To be fair one should say that the neutrino
scattering vs nucleon axial form factor is a kind of red herring; the
significance of nuclear effects is claimed {\it after} a fit of the
dipolar mass to the data is undertaken in which case astonishingly
precise values for the dipolar mass are inferred (see
e.g. \cite{Nie12} where extremely accurate values for $M_A$ are
quoted).  Since neutrino based determinations often imply certain and
some times questionable assumptions, it is instructive to review other
sources of information which at least do not rest on the same
assumptions.

On a fundamental level, {\it ab initio} calculations allow a direct
evaluation of the axial current matrix elements. The first lattice QCD
determination of the axial form factor \cite{Liu:1992ab} provided
$M_A=1.03(5) \,{\rm GeV}$ in agreement with world average neutrino data
at the time $M_A=1.032(36) \,{\rm GeV}$. However, subsequent
calculations~\cite{Capitani:1998ff} yield $M_A=1.5 \,{\rm GeV}$, a
number which has recently been confirmed~\cite{Alexandrou:2010hf} for
unphysical pion masses (about twice the physical value); the
corresponding dipolar axial mass is larger than the experimental one,
although there is some trend to agreement as the pion mass approaches
the physical value. The role of excited states has been analyzed in a
more recent lattice analysis~\cite{Junnarkar:2014jxa} confirming these
results.  In addition, Light cone QCD sum rules also overestimate the
experimental dipole fit by $30\%$\cite{Braun:2006hz} in the range $1 <
Q^2 < 4 \,{\rm GeV}^2$, a trend checked by subsequent
analyses~\cite{Wang:2006su,Erkol:2011qh} and agreeing also with
lattice calculations. While these QCD calculations are still subjected
to many improvements, one should also recognize that they generate a
family of axial nucleon form factors which fall within the
experimental band which is wide enough to pose again the pertinent
question on which are the correct ones within uncertainties. This
situation makes an interesting case of lifting the conventional fitting
strategy based on {\it ad hoc} parameterizations and incompatible data
in favor of assuming a theoretically founded axial nucleon form
factor with a credible uncertainty band generated by independent
fluctuations and not directly based on the neutrino-nucleus data under
discussion.  In this paper we propose a simple scheme furnishing these
requirements, see Section~\ref{sec:ffA}, and provide a framework where
the significance of different nuclear effects might be addressed.

On the more phenomenological hadronic level, the algebra of
fields~\cite{Lee:1967iu} which yields field-current
identities~\cite{Lee:1967iv} imply a generalized meson dominance which
has proven as a convenient tool to analyze many important hadronic
properties and most notably generalized vertex functions and hadronic
form factors~\cite{Frampton:1969ry}. In the particular case of
conserved currents, and more specifically axial-vector currents the
general form of the form factor is expected to be a sum of infinitely
many monopoles with isovector {\it axial} meson masses, whereas the
pQCD result~\cite{Carlson:1985zu} yields $G_A \sim 1/Q^4$. The
goodness of the Axial Vector Meson Dominance (AVMD) for the axial
nucleon form factor was posed in Ref.\cite{Gari:1984qs} by including
the strong vertex corrections. However, meson dominance implies
exchange of resonances which have a mass spectrum characterized by a
mass and a width. Amazingly there is a theoretical limit where
meson-dominance with narrow resonances is realized in QCD, namely the
large $N_c$-limit introduced by 'tHootf and Witten long
ago~\cite{'tHooft:1973jz,Witten:1979kh}; within the large $N_c$
expansion mesons become stable particles.  Their width-to-mass ratio
is $\Gamma_R/M_R= {\cal O} (1/N_c) \sim 0.33 $ which turns out to give
the correct order of magnitude of the average experimental value
0.12(8)~\cite{Masjuan:2012gc}. The phenomenological implications of
meson dominance within a large $N_c$ approach have been analyzed in
Ref.~\cite{Mas13} and, after natural uncertainty estimates based on
the resonance width, a good description of experimental data and
lattice results was achieved, with competitive accuracy.

Motivated by these theoretical insights, in the present paper we
explore the large-$N_c$-inspired parameterization of the nucleon axial
form factor \cite{Mas13}, see Section~\ref{sec:ffA}, 
%While some authors claim that, after including the relevant nuclear
%and reaction mechanisms, the MiniBooNE data are fully compatible with
%former determinations of the nucleon dipolar axial mass \cite{Nie12},
%the fact is that the dipole approximation cannot be justified from a
%field-theoretic point of view and is in contradiction with
%quark-hadron duality at large $N_c$ \cite{Mas13}. %
and explore the consequences of axial-vector dominance directly.  We
apply these findings to neutrino-nucleus scattering by starting with
the simplest nuclear model, i.e., the relativistic Fermi gas, see
Section~\ref{sec:qens}, which can be worked analytically.  We do this
without fitting any neutrino data in Section~\ref{sec:num}. This
simple approach allows to address more clearly some important issues
from a statistical point of view, and in particular to pin down the
region of $Q^2$-values where the axial nucleon form factor fits are
more sensitive to the existing neutrino-nucleus scattering data, see
Section~\ref{sec:goodness}. We finally summarize our results in
Section~\ref{sec:concl}.

\section{Axial-vector meson dominance and the axial form factor}
\label{sec:ffA}

\begin{figure}
\includegraphics[width= 8.5cm, bb=145 540 470 780]{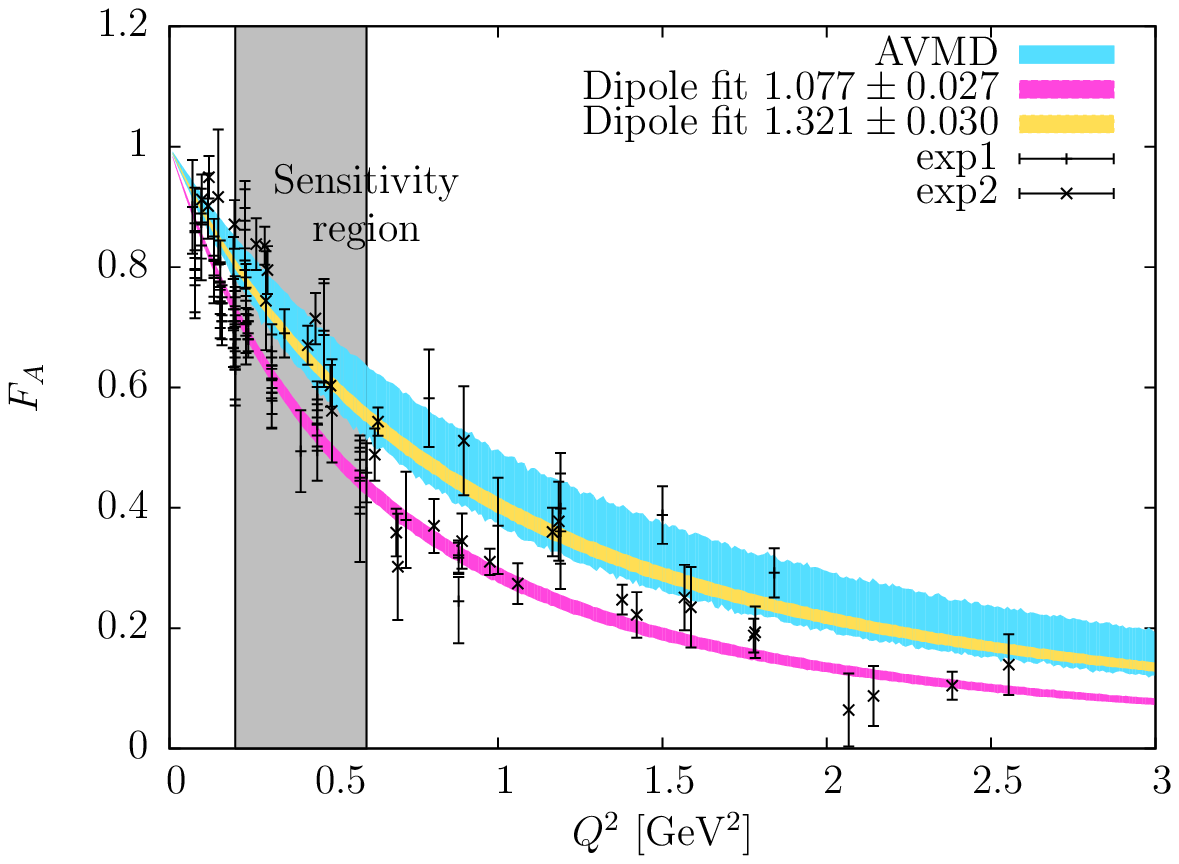}
\caption{ 
\label{fa1}
The axial meson dominance band prediction is compared with the
experimental data from the nucleon and with the dipole prediction bands
of ref. \cite{Nie12} using the Fermi-Gas approximation and some
relevant nuclear and reaction effects thereof. 
 Experimental data
are from refs. \cite{Nam70,Ben25,Fur70,Dom73,Joo76} (exp1) and
\cite{Bak81,Mil82,Kit83,Kit90} (exp2).
The axial form factor 
and the data are normalized to $F_A(0)=1$. 
 }
\end{figure}

\begin{figure}
\includegraphics[width= 8.5cm, bb=145 540 470 780]{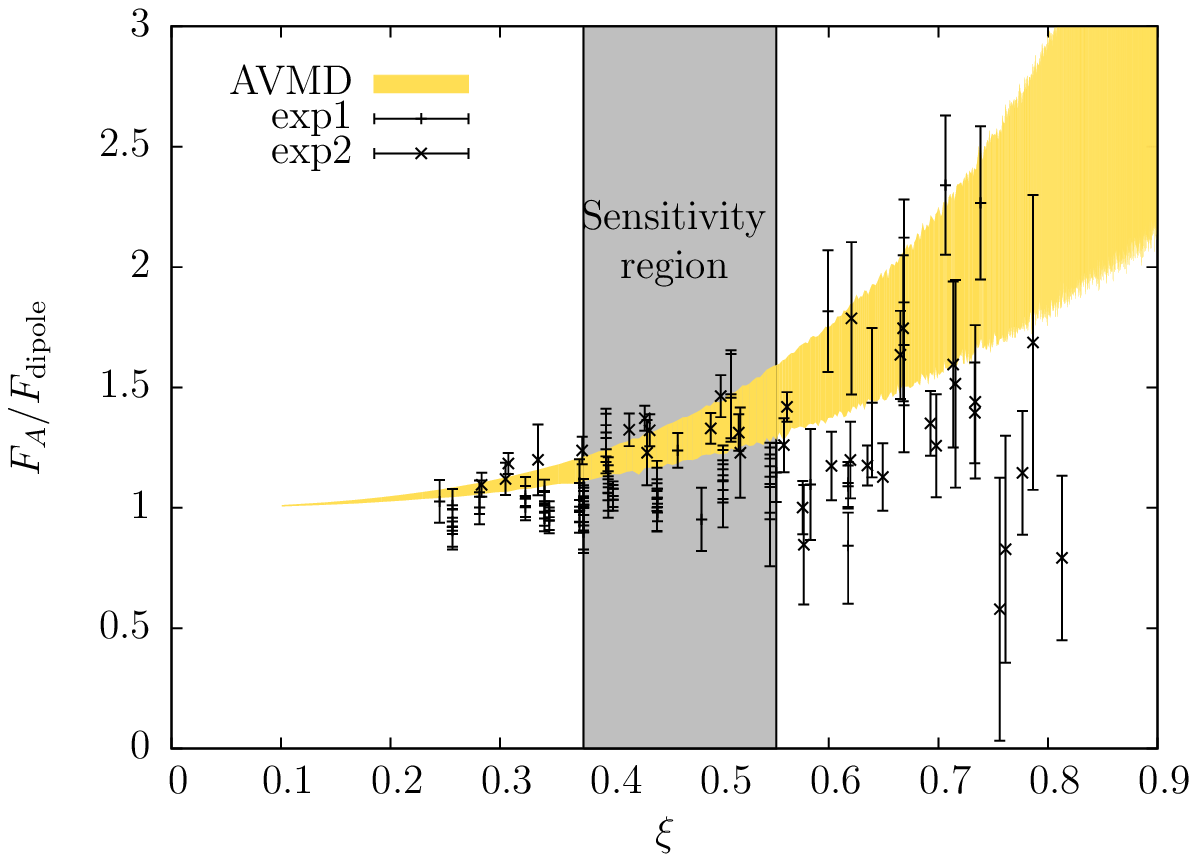}
\caption{ 
\label{fa2}
The axial meson dominance band prediction divided by the dipole
parameterization for $M_A=1.014$ MeV as a function of the dimensionless
variable $\xi=2/(1+\sqrt{1+4 m_N^2/Q^2})$ defined in
Ref.~\cite{Bodek:2007ym}.  Experimental data are from
refs. \cite{Nam70,Ben25,Fur70,Dom73,Joo76} (exp1) and
\cite{Bak81,Mil82,Kit83,Kit90} (exp2). The form factors and data are
normalized to one for $Q^2=0$. }
\end{figure}

Axial-Vector Meson Dominance (AVMD) was first introduced by Lee and
Zumino~\cite{Lee:1967iv} into particle physics as a very natural
generalization of the successful realization that Vector-Meson
dominance explained the bulk of electromagnetic form factors. It
simply states that the axial-vector current is given by the current
field identity, which for just $u,d$ quarks reads
\begin{eqnarray}
\vec J_A^\mu = \frac12 \bar q \gamma_\mu \gamma_5 \vec \tau q =  \sum_{A} f_A \partial^\nu \vec A_{\mu\nu}  + \sum_P f_P \partial_\mu \vec P \, , 
 \end{eqnarray}
where $f_A$ and $f_P$ are the decay amplitudes of the axial-vector
$A=a_1,a_1', \dots$ and pseudoscalar mesons $P=\pi,\pi', \dots$
respectively and $\vec A_{\mu\nu}= \partial_\mu \vec A_\nu 
-\partial_\nu \vec A_\mu $ is the corresponding field strength
tensor of the axial meson. This equation yields a generalized PCAC, which implies in turn
a
generalization~\cite{Dominguez:1971zg,Dominguez:1971hf,Dominguez:1976ut,Dominguez:1977nt,Dominguez:1977en,Dominguez:1984ka}
of the celebrated Goldberger-Treimann relation.

As a consequence the axial form factor of the nucleon can be written
as a sum of monopole form factors, 
\begin{eqnarray}
G_A(Q^2) = g_A \sum_{n} c_{n,a} \frac{m_{n,a}^2}{m_{n,a}^2+Q^2}
\end{eqnarray}
where $c_{n,a}=f_{n,a} g_{n,aNN} /g_A$ and $f_{n,a}$, $g_{n,aNN}$ and
$m_{n,a}$ are the vacuum amplitude, the coupling to the nucleon and
the mass respectively of the corresponding isovector-axial-vector
meson $n$. From $G_A(0)=g_A$ we have the normalization condition
\begin{eqnarray}
1 =  \sum_n c_{n,a} \, . 
\end{eqnarray}
The asymptotic pQCD result~\cite{Carlson:1985zu}, $G_A \sim 1/Q^4$, requires 
\begin{eqnarray}
0 =  \sum_n c_{n,a} m_{n,a}^2\, . 
\end{eqnarray}
In this paper we use the minimal hadronic ansatz for axial nucleon
form factor furnishing meson dominance and proper pQCD behavior
\begin{equation}
G_A(Q^2)
= g_A F_A(Q^2)
= g_A
\frac{m^2_{a_1}m^2_{a'_1}}{(m^2_{a_1}+Q^2)(m^2_{a'_1}+Q^2)}
\label{eq:ffA}
\end{equation}
with $g_A=1.267$, and where the axial meson masses are $m_{a_1}=1.230$
GeV, $m_{a'_1}=1.647$ GeV. As noted in Ref.~\cite{Mas13} one of the
problems with this ansatz is that generally the interpolating fields
are resonances which have a mass and a width, and we stand by the
solution proposed there to use the width as a genuine uncertainty of
the meson dominance ansatz. This generates a full band of predictions
which provide an uncertainty range for a formula of the form of
Eq.~(\ref{eq:ffA}). The experimental widths are
$\Gamma_{a_1}=0.425$ GeV, and $\Gamma_{a'_1}=0.254$ GeV as listed in
the PDG compilation~\cite{Beringer:1900zz}~\footnote{Of course this
  ansatz provides a value for the msr axial radius, $\langle r^2
  \rangle_A = 6/m_{a_1}^2+ 6/m_{a_1'}^2 $.  One can add a further
  axial state fixing the radius to its precise value, and comply to
  the pQCD short distance constraint but the effect is not large. This
  way one might take into account, the tiny and predictable
  differences between axial radii determined by either
  electroproduction or neutrino scattering.}.  The masses are only the
central values of the axial mesons spectra.  We use the half-width
rule to generate random values for $m_{a_1}$ and $m_{a'_1}$ following
Gaussian distributions with variances $\Gamma_{a_1}/2$ and
$\Gamma_{a_1'}/2$ respectively. This provides a distribution band for
the axial form factor~\cite{Mas13} which is slightly above the bulk of
the abundant and incompatible $G_A$ data, see Fig.~\ref{fa1}, but
agrees well with the
lattice~~\cite{Capitani:1998ff,Alexandrou:2010hf,Junnarkar:2014jxa}
and light cone QCD sum rules
estimates~\cite{Braun:2006hz,Wang:2006su,Erkol:2011qh}.

It might be useful to provide a parameterization of the uncertainty
bands of the AVMD form factor as $F_A^{\rm lower}(Q^2)\leq
F_A(Q^2)\leq F_A^{\rm upper}(Q^2)$.  The lower and upper axial form
factors are defined as the boundaries of the usual $1\sigma$ 68\%
confidence level region.  They can be parameterized as a product of
two monopoles, similarly to Eq. (6), as
\begin{equation}
F_A^{\alpha}(Q^2)= 
\frac{\Lambda^2_{\alpha} \Lambda^2_{\alpha'}}
{(\Lambda^2_{\alpha}+Q^2)(\Lambda^2_{\alpha'}+Q^2)}
\end{equation}
with $\alpha=\rm upper\,,lower$. By a fit in the range 
$0 \le Q^2 \le  3 \,\rm GeV^2$, 
corresponding to the blue band of Fig.1, the
cut-off parameters turn out to be 
$(\Lambda_{\rm lower},\Lambda_{\rm lower'})=(0.97486,1.73345)$, 
and 
$(\Lambda_{\rm upper},\Lambda_{\rm upper'}) = (1.5436,1.54194)$, 
respectively (in  GeV) .  
These values illustrate the fact that the fluctuations of
  the form factor do not necessarily correspond to a single dipole
  mass fluctuation.

We will thus apply this axial form factor band for the neutrino cross
section theoretical predictions. Our point of view is that given the many
effects which might contribute to neutrino-nucleus scattering it may
be sensible to use a credible form factor with an error estimate based
on a different source of data, without resting on a specific fit to
the neutrino data. It was found in Ref.~\cite{Mas13} that the
large-$N_c$ meson dominated form factors with pQCD constraints and
supplemented with the half-width rule for an uncertainty estimate
worked well also for other form factors such as electromagnetic,
scalar and gravitational form factors. As a general rule uncertainties
turned out to be comparable or smaller than lattice QCD predictions
but larger than experimental data.

After presenting our main results, for completeness we will also analyze the
conventional approach of fitting Eq.~(\ref{eq:ffA}) 
to the MiniBoone data. We want to investigate the traditional point of
view of {\it assuming} certain nuclear effects before undertaking a
fit of the axial form factor of the nucleon.  For example in
Ref.~\cite{Nie12} a model with nucleon spectral functions, RPA
correlations and MEC has been considered and a fit to the axial form
factor has been undertaken assuming a fixed $\Delta-N$-transition form
factor. We want to understand why in these studies one can extract
more accurate information on the axial form factor than on the nuclear
model response functions.

Anticipating some of the results to be discussed below, and for a
comparison we depict also in Fig. \ref{fa1} the results found in the
analysis of ~\cite{Nie12} where the role of nuclear effects beyond the
local Fermi-Gas have been addressed when a dipole form factor is
fitted to the neutrino scattering data. As can be deduced from
Fig.~\ref{fa1} the statistical errors are comparable when including
the additional nuclear effects and the large and quite visible
systematic change between the two fits is comparable to the spread
generated by our AVMD form factor. Following the scheme of
Ref.~\cite{Bodek:2007ym} we also plot in Fig.~\ref{fa2} the ratio
between the AVMD form factor and the dipole form factor,
Eq.~(\ref{eq:dipole}), with $M_A=1.014 \,{\rm GeV}$ in terms of the
dimensionless variable $\xi$ defined there.  This quotient was fitted
in ~\cite{Bodek:2007ym} by including an interpolating polynomial
without discarding any of the compiled data which, as we have
mentioned, are incompatible as a whole~\footnote{If would be
  interesting to check if the rather small uncertainties obtained in
  Ref.~\cite{Bodek:2007ym} are triggered by the inevitable large
  $\chi^2$ values which are usually obtained when fitting mutually
  incompatible data and by the stiffness against fitting parameter
  variations.  Unfortunately, no $\chi^2$ value has been quoted and it
  is difficult to asses the goodness of fit.}. As we can see the AVMD
model produces a spread compatible with the spread of the region
covered by the form factor data.

In both Fig.~\ref{fa1} and Fig.~\ref{fa2} we also highlight in shaded
gray the main $Q^2$ region where fluctuations in the axial form factor
have a sizable impact in the MiniBooNE data, as will be discussed
below in Section~\ref{sec:goodness}. Thus, our AVMD motivated axial
form factor describes reasonably well the known data spread in the
$Q^2$-region relevant for the MiniBooNE experiment.

\section{Quasielastic neutrino scattering}
\label{sec:qens}

In this paper we are interested in the charged-current quasielastic
(CCQE) reactions in nuclei induced by neutrinos.  In particular we
compute the $(\nu_\mu,\mu^-)$ cross section.  The total energies of
the incident neutrino and detected muon are $\epsilon=E_\nu$,
$\epsilon'=m_\mu+T_\mu$, and their momenta are $\nk,\nk'$.  The
four-momentum transfer is $k^\mu-k'{}^\mu=(\omega,\nq)$, with
$Q^2=q^2-\omega^2 > 0$.
% We
%use a coordinate system with the $z$-axis pointing along $\nq$ and the
%$x$-axis along the transverse component of the incident neutrino. 
If
the lepton scattering angle is $\theta$, the double-differential cross
section can be written as \cite{Ama05a,Ama05b}
\begin{equation}
\frac{d^2\sigma}{dT_\mu d\cos\theta}(E_\nu)=
\left(\frac{M_W^2}{M_W^2+Q^2}\right)^2\frac{G^2\cos^2\theta_c}{4\pi}
\frac{k'}{\epsilon}v_0  S_{\pm}
\end{equation}
Here $G=1.166\times 10^{-11}\quad\rm MeV^{-2} \sim 10^{-5}/ m_p^2$ is
the Fermi constant, $\theta_c$ is the Cabibbo angle,
$\cos\theta_c=0.975$, and the kinematical factor $v_0=
(\epsilon+\epsilon')^2-q^2$.

The nuclear structure function $S_{\pm}$
is defined as a linear combination of the five nuclear response functions 
 (+ is for neutrinos and $-$ is for antineutrinos)
\begin{equation}
S_{\pm}=
V_{CC} R_{CC}+
2{V}_{CL} R_{CL}+
{V}_{LL} R_{LL}+
{V}_{T} R_{T}
\pm
2{V}_{T'} R_{T'}\ ,
\end{equation}
where the $V_K$ coefficients depends only on the neutrino and muon kinematics 
and do not depend on the details of the nuclear target.
\begin{eqnarray}
{V}_{CC}
&=&
1-\delta^2\frac{Q^2}{v_0}
\label{vcc}\\
{V}_{CL}
&=&
\frac{\omega}{q}+\frac{\delta^2}{\rho'}
\frac{Q^2}{v_0}
\\
{V}_{LL}
&=&
\frac{\omega^2}{q^2}+
\left(1+\frac{2\omega}{q\rho'}+\rho\delta^2\right)\delta^2
\frac{Q^2}{v_0}
\\
{V}_{T}
&=&
\frac{Q^2}{v_0}
+\frac{\rho}{2}-
\frac{\delta^2}{\rho'}
\left(\frac{\omega}{q}+\frac12\rho\rho'\delta^2\right)
\frac{Q^2}{v_0}
\\
{V}_{T'}
&=&
\frac{1}{\rho'}
\left(1-\frac{\omega\rho'}{q}\delta^2\right)
\frac{Q^2}{v_0}.
\label{vtp}
\end{eqnarray}
where  we have defined the dimensionless factors
$\delta = m'/\sqrt{Q^2}$, proportional to the muon mass $m'$, 
$\rho = Q^2/q^2$, and $\rho' = q/(\epsilon+\epsilon')$.

We evaluate the five nuclear response functions $R_K$, $K=CC, CL, LL,
T, T'$ ($C$=Coulomb, $L$=longitudinal, $T$=transverse). following the
simplest approach that treats exactly relativity, gauge invariance and
translational invariance, that is the relativistic Fermi gas model
(RFG) \cite{Ama05a,Ama05b}.  The single nucleons are described by
plane wave spinors and the response functions are analytical.  It is a
remarkable result that the nuclear response function $R_K$ of the RFG
is proportional to a single-nucleon response function $U_K$ times the
so-called scaling function $f(\psi)$
\begin{equation} \label{rfg}
R_K = \frac{N \xi_F}{m_N \eta_F^3 \kappa}  U_K  f(\psi)
\end{equation}
where $N$ is the neutron number, $\eta_F=k_F/m_N$, and
 $\xi_F=\sqrt{1+\eta_F^2}-1$.
The scaling function is defined as
\begin{equation}
f(\psi)=\frac34 (1-\psi^2)\theta(1-\psi^2)
\end{equation}
where $\theta $ is the Heavyside step function and $\psi$ is the
scaling variable
\begin{equation}
\psi^2 = \frac{1}{\xi_F} {\rm max}
\left\{
\kappa\sqrt{1+\frac{1}{\tau}}-\lambda -1 , \xi_F-2\lambda
\right\}
\end{equation}
where $\lambda=\omega/(2m_N)$, $\kappa=q/(2m_N)$, and $\tau=\kappa^2-\lambda^2$.

Finally, we give the single-nucleon responses $U_K$.
For $K=CC$ it is the sum of vector 
and axial-vector response, in turns written as  the sum of conserved (c.)
plus non conserved (n.c.) parts,
\begin{eqnarray}
U_{CC} &=& U_{CC}^V+
\left(U_{CC}^A\right)_{\rm c.}
+\left(U_{CC}^A\right)_{\rm n.c.}
\end{eqnarray}
For the vector CC response we have
\begin{eqnarray}
U_{CC}^V &=&
\frac{\kappa^2}{\tau}
\left[ (2G_E^V)^2+\frac{(2G_E^V)^2+\tau (2G_M^V)^2}{1+\tau}\Delta
\right]\ ,
\end{eqnarray}
where $G_E^V$ and $G_M^V$ are the isovector electric and magnetic
nucleon form factors (we use Galster's parameterization), and
\begin{equation}
\Delta= \frac{\tau}{\kappa^ 2}\xi_F(1-\psi^2)
\left[\kappa\sqrt{1+\frac{1}{\tau}}+\frac{\xi_F}{3}(1-\psi^2)\right].
\end{equation}
The axial-vector CC response is  the sum of conserved (c.)
plus non conserved (n.c.) parts,
\begin{eqnarray}
\left(U_{CC}^A\right)_{\rm c.}
&=&
\frac{\kappa^2}{\tau}G_A^2\Delta
\\
\left(U_{CC}^A\right)_{\rm n.c.}
&=&
\frac{\lambda^2}{\tau}(G_A - \tau G_P )^2.
\end{eqnarray}
where $G_A$ is the nucleon axial-vector form factor and 
 $G_P$ the pseudoscalar axial form factor.
From PCAC the pseudoscalar form factor is
\begin{equation}
G_P =  \frac{4m_N^2}{m_\pi^2+Q^2}G_A.
\end{equation}

Similarly, for $K=CL,LL$ we have
\begin{eqnarray}
U_{CL} &=& U_{CL}^V+\left(U_{CL}^A\right)_{\rm c.}
+\left(U_{CL}^A\right)_{\rm n.c.}
\\
U_{LL} &=& U_{LL}^V+\left(U_{LL}^A\right)_{\rm c.}
+\left(U_{LL}^A\right)_{\rm n.c.}\ ,
\end{eqnarray}
The vector and conserved axial-vector parts are determined by
current conservation
\begin{eqnarray}
U_{CL}^V &=& -\frac{\lambda}{\kappa}U_{CC}^V
\\
\left(U_{CL}^A\right)_{\rm c.}
&=& -\frac{\lambda}{\kappa}
\left(U_{CC}^A\right)_{\rm c.}
\\
U_{LL}^V &=& \frac{\lambda^2}{\kappa^2}U_{CC}^V
\\
\left(U_{LL}^A\right)_{\rm c.}
&=& \frac{\lambda^2}{\kappa^2}
\left(U_{CC}^A\right)_{\rm c.}\ ,
\end{eqnarray}
while the n.c. parts are
\begin{eqnarray}
\left(U_{CL}^A\right)_{\rm n.c.}
&=& -\frac{\lambda\kappa}{\tau}(G_A - \tau G_P )^2
\\
\left(U_{LL}^A\right)_{\rm n.c.}
&=& \frac{\kappa^2}{\tau}(G_A - \tau G_P )^2\ .
\end{eqnarray}
Finally the transverse responses are given by
\begin{eqnarray}
U_T &=& U_T^V+U_T^A
\\
U_T^V &=&  2\tau(2G_M^V)^2+\frac{(2G_E^V)^2+\tau (2G_M^V)^2}{1+\tau}\Delta
\\
U_T^A &=& 2(1+\tau)G_A^2 + G_A^2 \Delta
\\
U_{T'} &=& 2G_A(2G_M^V) \sqrt{\tau(1+\tau)}[1+\tilde{\Delta}]
\end{eqnarray}
with
\begin{equation}
\tilde{\Delta}=\sqrt{\frac{\tau}{1+\tau}}\frac{\xi_F(1-\psi^2)}{2\kappa}\ .
\end{equation}

\section{Numerical results}
\label{sec:num}

\begin{figure}
\includegraphics[width= 8.5cm, bb=145 540 470 780]{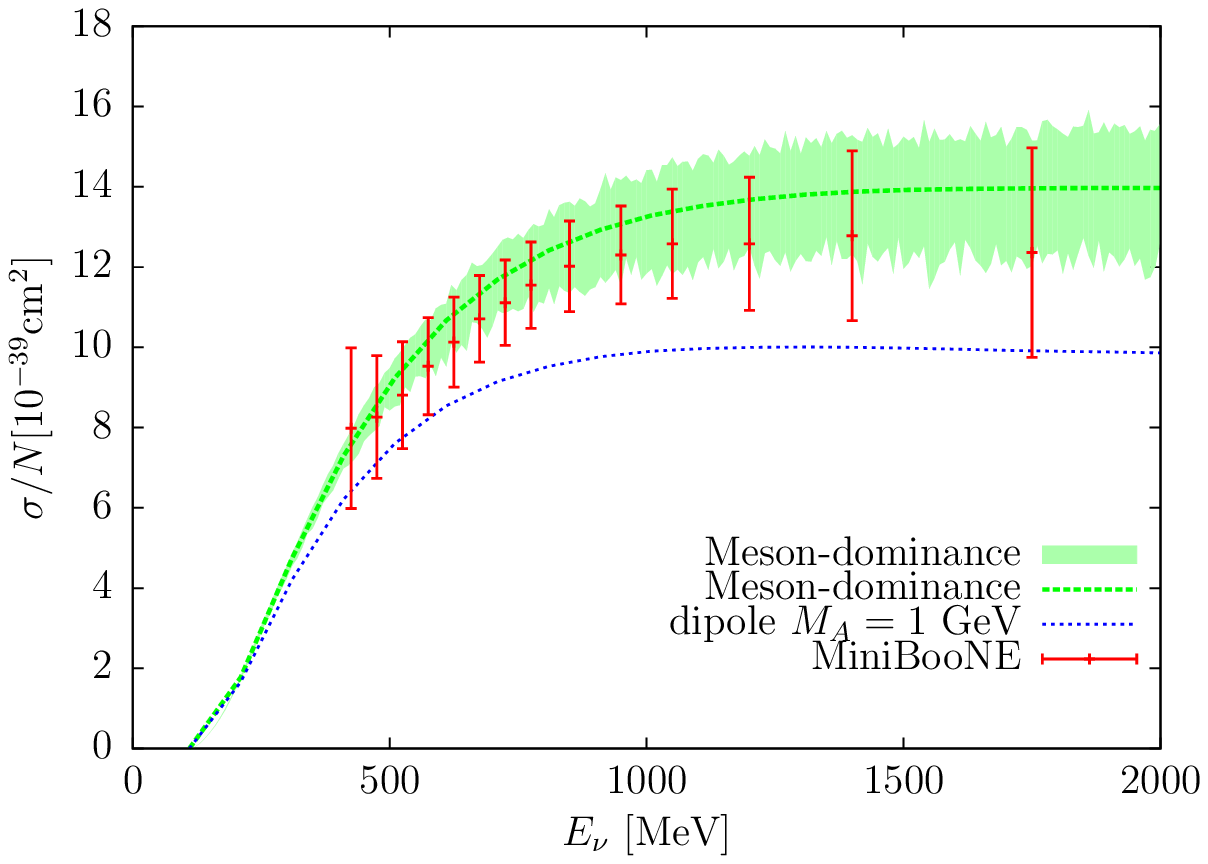}
\caption{ 
\label{cros}
Integrated quasielastic neutrino cross section of $^{12}$C.
  The axial meson dominance band prediction is centered around the
  axial meson masses, and it is compared to the dipole parameterization with
  dipolar axial mass $M_A=1$ GeV.  The experimental data are from MiniBooNE experiment.
}
\end{figure}

In Fig.\ref{cros} we show the AVMD predictions for the
total integrated CCQE cross section
\begin{equation}
\sigma(E_\nu) = 
\int dT_{\mu} \int d\cos\theta 
\frac{d^2\sigma}{dT_\mu d\cos\theta}(E_\nu).
\end{equation}
The theoretical uncertainties represented by the displayed band have
been computed by a Monte Carlo calculation assuming a Gaussian
distribution for the axial meson mass distributions.  For comparison
we show also the results obtained with a dipole axial form factor with
$M_A=1$ GeV. The MiniBooNE data are compatible with the axial
meson-dominance predictions. Note that no attempts to fit the
experimental data have been made.  The only parameter of the RFG model
is the Fermi momentum $k_F=225$ MeV.

The MiniBooNE unfolded energy dependent cross section is model
dependent based on a reconstruction of the neutrino energies assuming
a quasielastic interaction with a neutron at rest. These data suffer
from uncertainties driven by the model dependence of the neutrino
energy reconstruction. For proper and useful comparisons, the
flux-averaged doubly differential cross section should be used. We
compute this cross section as
\begin{equation}
\frac{d^2\sigma}{dT_\mu d\cos\theta} 
= \frac{
\int dE_\nu \phi(E_\nu)
\frac{d^2\sigma}{dT_\mu d\cos\theta}(E_\nu)
} 
{\int dE_\nu \phi(E_\nu)} 
\end{equation}
where $\phi(E_\mu)$ is the incident neutrino flux.

\begin{figure}[ht]
\includegraphics[width= 8cm, bb=50 200 500 780]{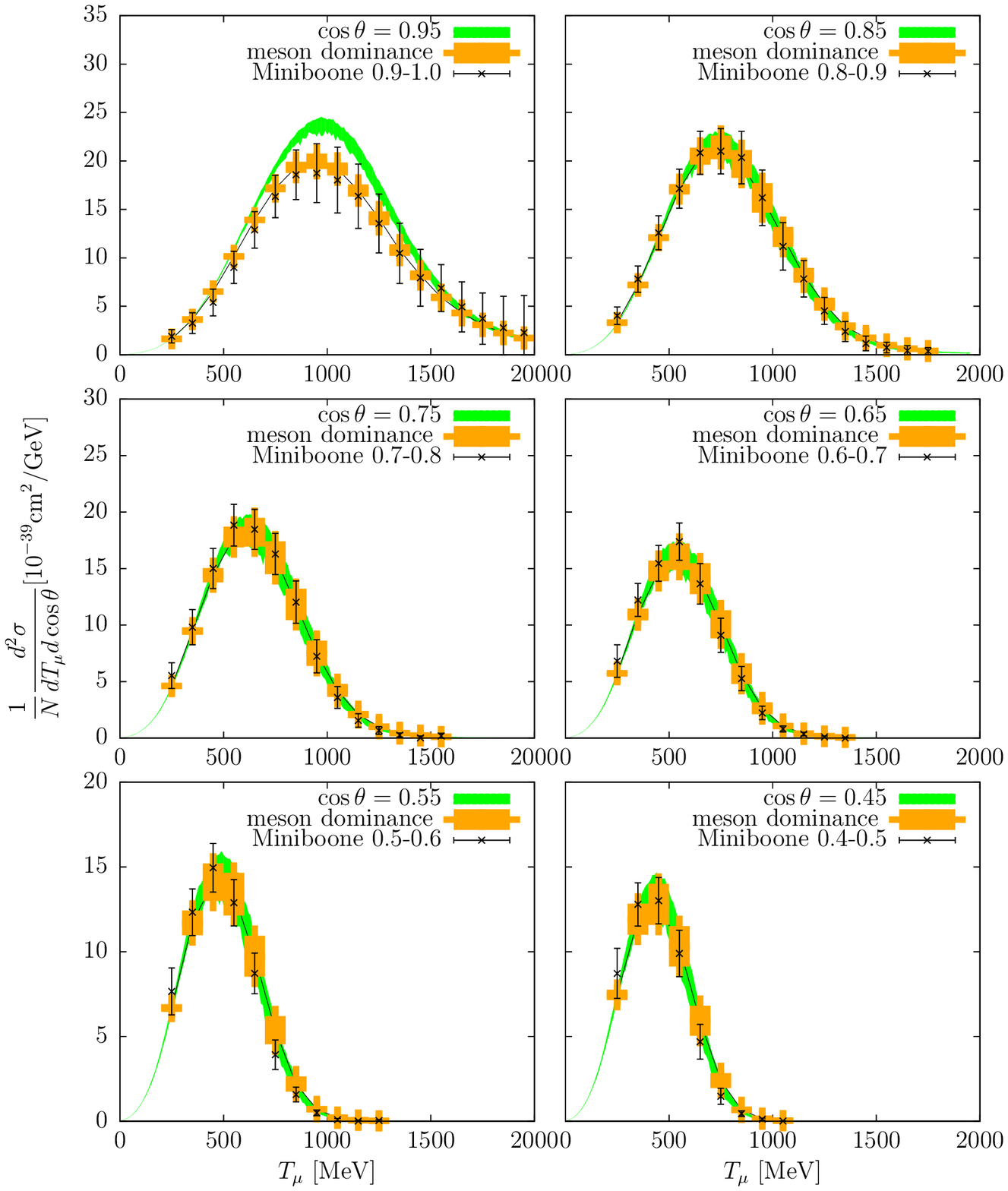}
\caption{ 
\label{dcros}
Flux-averaged doubly differential CCQE cross section as a
  function of the muon kinetic energy. The continuous band predictions
  (green) have been computed for fixed values of $\cos\theta$ at the
  center of the experimental bins. The discrete meson dominance
  predictions have been computed by integrating the
  doubly-differential cross section over each discrete bin.  }
\end{figure}

In figure \ref{dcros} we show results for the flux-averaged doubly
differential CCQE cross section as a function of the muon kinetic
energy. The bands are the axial meson-dominance model predictions for
fixed values of $\cos\theta$ at the center of the experimental
bins.  

\begin{figure}
\includegraphics[width= 8.5cm, bb=145 540 470 780]{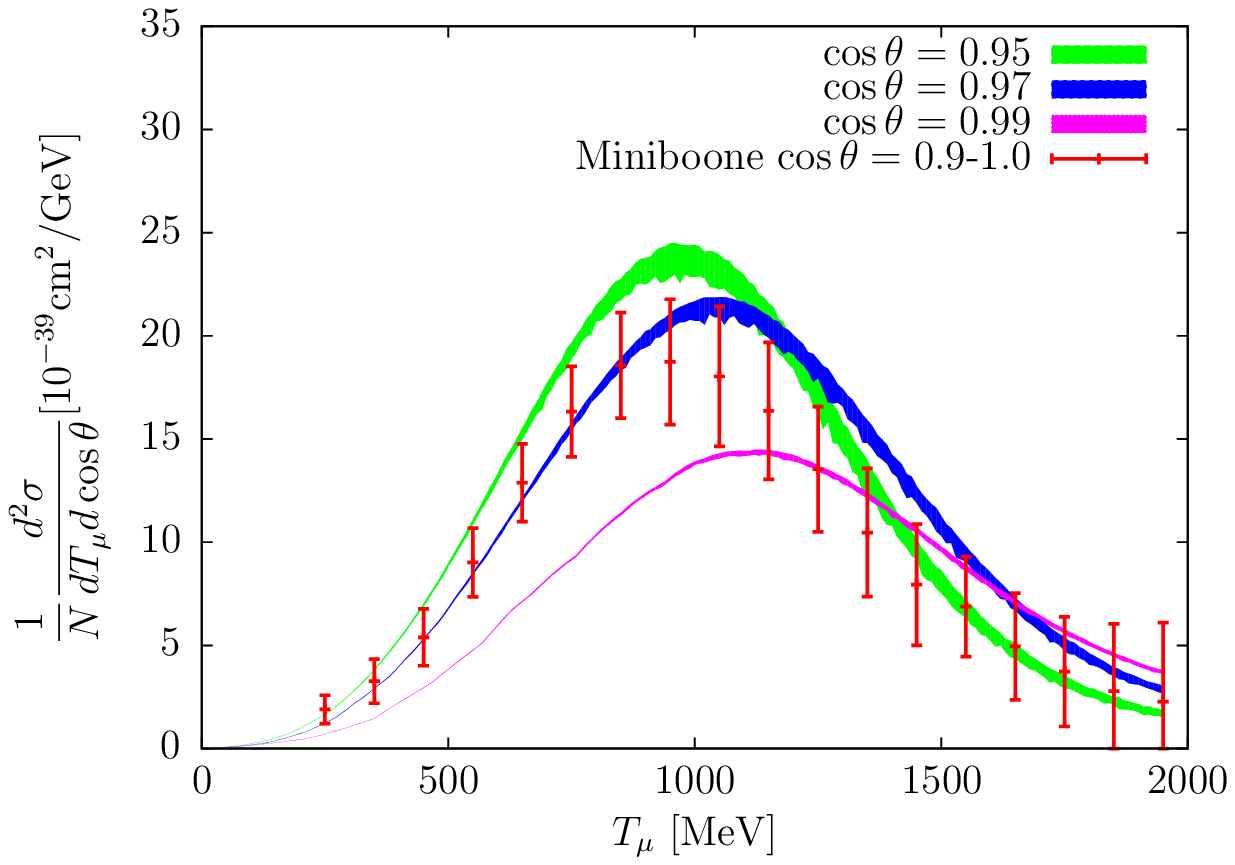}
\caption{ 
\label{angulos}
Flux-averaged doubly differential CCQE cross section
 for several values of $\cos\theta$, 
and for low scattering angles,
compared with the experimental data for the bin 
$\cos\theta=0.9$-1.0. 
}
\end{figure}

The MiniBooNE $\nu_{\mu}$ CCQE flux-integrated double differential
cross section is provided in bins $(t_i,t_{i+1})$ of $T_{\mu}$ and bins
$(c_j,c_{j+1})$ of $\cos\theta$. 
The size of the bins is $c_{j+1}-c_j=\Delta c = \Delta\cos\theta_{\mu}=0.1$, 
and $t_{i+1}-t_i= \Delta t = \Delta T_{\mu}=0.1$ GeV. 

For a meaningful comparison with the experimental data we have computed
the averaged cross section for each bin, 
by integrating the doubly-differential cross section over each
discrete bin.
\begin{equation} \label{average}
\Sigma_{ij} = \frac{1}{\Delta t \Delta c}
\int_{t_i}^{t_{i+1}} dT_{\mu} \int_{c_{j}}^{c_{j+1}} d\cos\theta 
\frac{d^2\sigma}{dT_\mu d\cos\theta}.
\end{equation}

\begin{figure}[ht]
\includegraphics[width= 9.5cm]{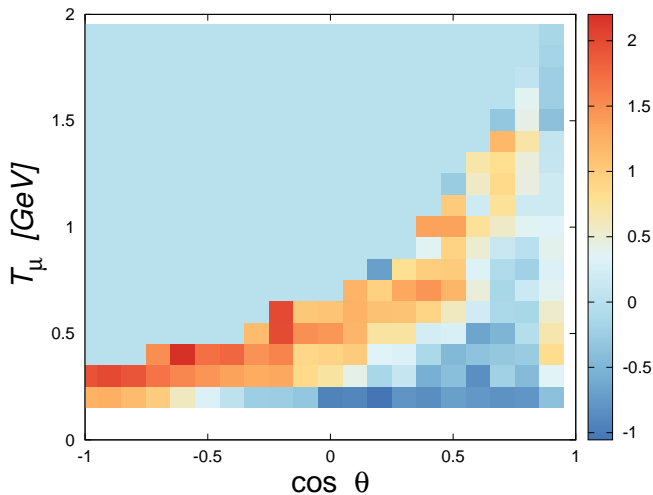}
\caption{
\label{chi2}
The $\chi_{ij}$ computed values  
for each  bin pair are
 represented in the $(\cos\theta_{\mu},T_{\mu})$ plane 
as a color image.
 }
\end{figure}

The axial vector dominance predictions for the averaged cross section
$\Sigma_{ij}$ are also shown in figure \ref{dcros}, where they are
compared to the experimental data. The theoretical errors are again
computed by assuming a Gaussian distribution of the axial meson
masses. Note that the averaged cross section for low scattering
angles, bin $\cos\theta=0.9$-1.0, is quite different from the cross
section at the central value $\cos\theta=0.95$. This is due to the
strong angular dependence of the differential cross section for small
angles, as can be seen in Fig. \ref{angulos}. Therefore the
integration of the cross section over the bin is crucial to get the
correct average. Note also that in this region the momentum transfer
takes the smallest values compatible with energy transfer, and one
expects that the model dependence of the results be maximized.  As a
matter of fact according to Ref.~\cite{Amaro:2006tf} the shell
structure effects for both discrete and continuum are essentially washed
out in favor of the RFG. For larger scattering angles, the angular
dependence is mild, and the value of the cross section at the center
of the bin is closer to the average, Eq. (\ref{average}), as can be
seen in Fig. \ref{dcros}.

\section{Goodness of the model}
\label{sec:goodness}

To get a global measure of the goodness of the theoretical model in
describing the experimental data requires including both theoretical
and experimental uncertainties. We thus compute the distance of theory
to data as given by a $\chi^2$ metric, defined as $\chi^2 = \sum_{i,j}
\chi_{ij}^2$.  The $\chi_{ij}$ matrix provides the distance between
theory and experiment within each bin $(i,j)$, in units of the total
uncertainty.  It is defined as
\begin{equation}\label{chimatrix}
\chi_{ij} = \frac{\Sigma_{ij}^{(th)}-\Sigma_{ij}^{(exp)}}
{\sqrt{(\Delta\Sigma_{ij}^{(th)})^2 +(\Delta\Sigma_{ij}^{(exp)})^2}} 
\end{equation}
where $\Delta\Sigma_{ij}^{(exp)}$ is the experimental error.  and
$\Delta\Sigma_{ij}^{(th)}$ is the theoretical uncertainty due to the
physical widths of the axial mesons. In Fig.  \ref{chi2} we show the
matrix values $\chi_{ij}$ computed for all the bins of the MiniBooNE
CCQE neutrino experiment.  We obtain $\chi^2= 111$, so that dividing
by the number of bins $N=137$, we get $ \chi^2/N = 0.81$.  Globally
the model agrees remarkably well with data taking into account that we
do not minimize any $\chi^2$ and we just compute it.

While the $\chi^2$ value seems to be acceptable, let us analyze the
assumptions underlying the comparison and its statistical significance
in some more detail. We are just testing that the difference between
the theory and the data should behave as a random variable, namely a
standardized normal distribution. However, a look to the
Fig.~\ref{chi2} reveals that the level of disagreement is located at
the edges of the plot, while we should expect a more uniform pattern
globally if the $\chi_{ij}$ were distributed randomly. This can be
further elucidated by analyzing the differences. We find that there is
a strong asymmetry in the residuals $\chi_{ij}$, indicating gross
systematic differences. Thus, we believe that these large
discrepancies are possibly beyond the applicability of the RFG. At the
same time one should also admit that the double binning procedures,
essential for a proper comparison with the data, tend to wash out
nuclear effects.

As we anticipated in our discussion around Fig.~\ref{fa1} some fits to
the dipolar axial mass do generate rather good $\chi^2$ values and
unprecedented accuracy for the axial form factor~\cite{Nie12}. Let us
remind that, a too low value is as bad as a too high value, since the
$\chi^2$-distribution for a large number of degrees of freedom $\nu =
N-P$ behaves as a Gaussian distribution and thus it must be
$\chi^2/\nu=1 \pm \sqrt{2/\nu}$ within $1\sigma$ confidence level. For
instance in Ref.~\cite{Nie12} a value of $\chi^2/\nu =
33/(137-2)=0.24$ was obtained which is outside the expected confidence
level by $6 \sigma$. This suggest that experimental errors may be too
large, and the question is whether errors can be reduced without
destroying the Gaussian nature of the fluctuations. Moreover, let us
remind that the statistical approach based on $\chi^2$-fits deals with
testing the validity of a given functional form for the {\it true}
form factor, while despite the much extended popularity there is no
field theoretical support for a dipole form factor.

In order to understand those results we have performed a conventional
$\chi^2$-fit with two axial masses as minimization parameters.  For
this fit we include only the experimental errors in the denominator of
Eq. (\ref{chimatrix}).  As in Ref. \cite{Nie12} we normalize the data by a factor
$\lambda=0.96$ and subtract a constant $Q$ -value $Q_b=17$ MeV to the
energies of the particle-hole excitations (note that while this
modification by hand of the RFG energies improves the fit, the gauge
invariance of the model is broken).  We find the minimum at $m_{a_1}=
m_{a_1'}= 1293$ with $\chi^2/\nu=0.31$. We have tested the normality
of residuals, and we find that they very likely correspond to a
Gaussian distribution. This indicates that the experimental errors
should probably be re-scaled by a factor less than $1/2$, i.e., the
fit would be acceptable if the errors were twice smaller than stated
in the experiment. This observation concerns all previous
determinations of the dipolar axial mass from these neutrino data,
based on fits trying to minimize the discrepancies with the
experiment. 
Note that we {\it are not} disputing the existence of
certain well known important nuclear effects. The total uncertainty on
the theoretical neutrino-nucleus cross section can be due to
uncertainties on both the nuclear effects and on the axial form
factor.  Here we focus on the size of the axial form factor
uncertainties since they are obviously not small.

\begin{figure}[ht]
\includegraphics[width=9cm]{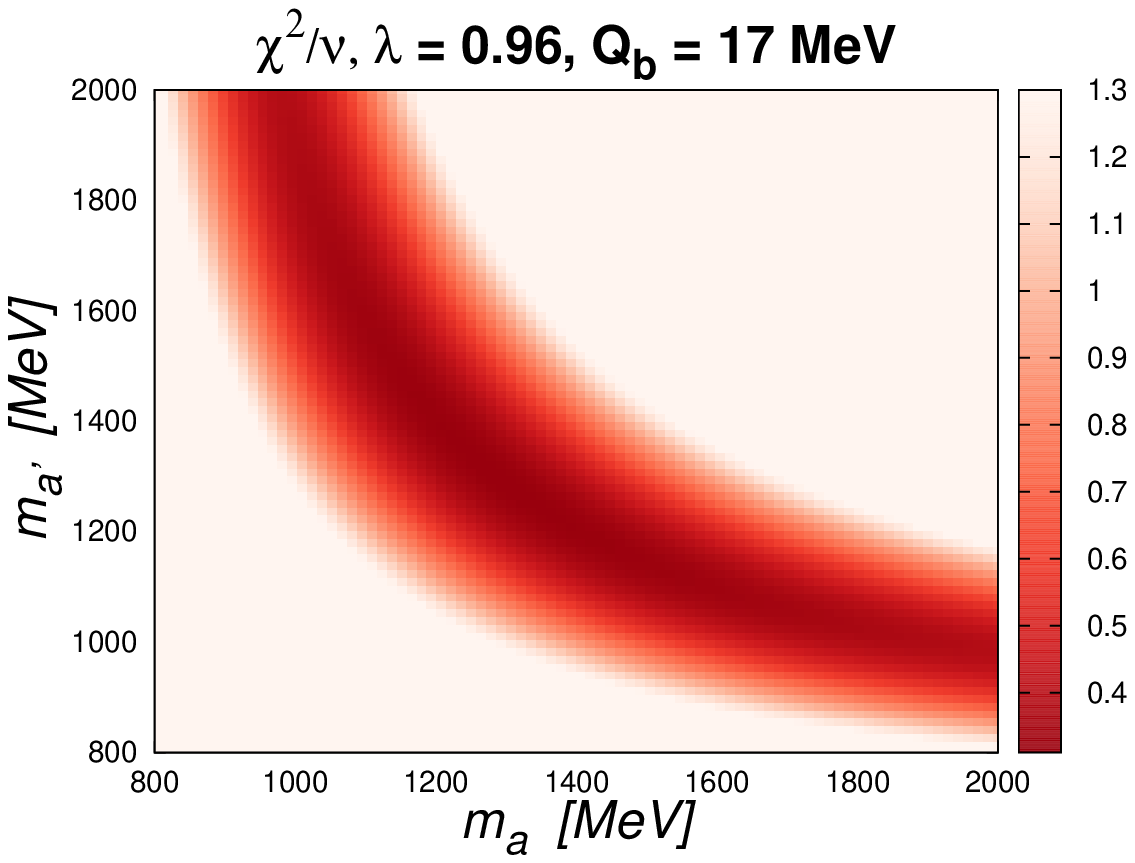}
\includegraphics[width=9cm]{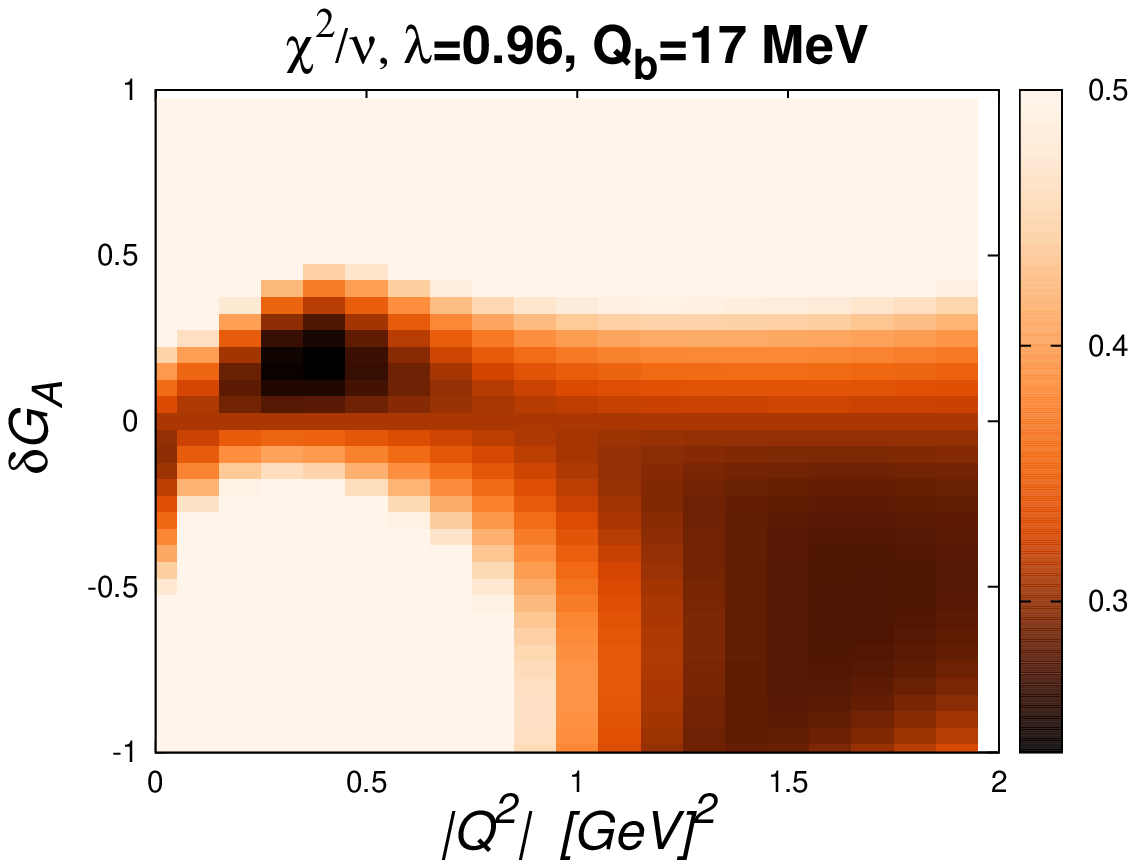}
\caption{ 
\label{fig:fit-errors}
Top panel: $\chi^2$ contour plot for the fitted axial masses.  Bottom
panel: $\chi^2$ contour plots for the errors in the axial form factor
$\delta G_A(Q^2)$ when the $Q^2$ values are binned with $\Delta Q^2 =
0.1 \,{\rm GeV}^2$ in the range $0.1 \,{\rm GeV}^2 \le Q^2 \le 2 \,{\rm
  GeV}^2 $.}
\end{figure}

In Fig.~\ref{fig:fit-errors} 
we plot the $\chi^2/\nu$ values as a
function of the two axial masses, showing that they are highly
correlated.  While the dipole form factor (two equal axial masses) is
contained in the confidence region around the minimum, it is not the
only allowed solution as two different axial masses also provide
acceptable fits.

In order to study the sensibility of the results against general
variations of the form factor, we also show in
fig.~\ref{fig:fit-errors} the $\chi^2/\nu$ contour plots for the
errors in the axial form factor $\delta G_A(Q^2)$ when the $Q^2$
values are binned with $\Delta Q^2 = 0.1 \,{\rm GeV}^2$ in the range
$Q^2 \le 2 {\rm\, GeV}^2 $.  The $\chi^2$ for each value of $\delta
G_A$ in a $Q^2$ bin has been computed by adding the specified value of
$\delta G_A$ to the form factor at the $Q^2$ values of the
corresponding bin only.  This shows that the value of $\chi^2$ can be
lowered further for more general variations of the axial form
factor. To explore this issue deeper we have performed simultaneous
variations of $\delta G_A(Q^2)$ in twenty $Q^2$ bins.  A new minimum was
found giving $\chi^2=23.8$ and $\chi^2/\nu = 0.17$.  As seen in
fig. \ref{fig:fit-errors} ---and verified by our minimization--- the
data seems to favor a larger form factor around $Q^2=0.4$ GeV$^2$
and a smaller one around 1.2 -- 2 GeV$^2$.

Note that, while our analysis here is focused on the quasielastic
neutrino-nucleus scattering, similar meson-dominance ideas for the nucleon and
$\Delta$-resonance have been discussed in previous work by Dominguez
and collaborators~\cite{Dominguez:2004bx,Dominguez:2007hg} suggesting
a possible extension to $N-\Delta$ transition form factors.

\section{Conclusions and outlook}
\label{sec:concl}

Most of the previous analyzes of the axial form factor assuming a
dipolar form and fitting MiniBoone neutrino-nucleus scattering data
provide unprecedented accurate but incompatible determinations of the
dipolar axial mass, regardless of the assumed nuclear model. This
suggests that while the proposed dipolar parameterization minimizes
the mean squared distance between theory and data, it does not account
properly for the experimental data fluctuations, introducing a
systematic bias and invalidating the conventional least squares
fitting strategy assumptions. Besides, the large spread of the many
experimental data for the axial form factor is not sharpened by the
currently available lattice QCD calculations or QCD sum rules
estimates, where a direct determination of the axial current matrix
element has been undertaken. As a consequence, the validation of known
nuclear effects in neutrino-nucleus scattering is hampered by the many
contradicting determinations of the axial form factor already in the
quasielastic region.

We take here a different perspective admitting from the beginning the
existence of an uncertainty band in the axial form factor. We assume a
theoretically based axial form factor with an a priori uncertainty
estimate, regardless of the neutrino data we intend to
describe. Namely, we use the minimal AVMD compatible with pQCD, an
ansatz motivated by quite general large $N_c$ features and which
requires just the $a_1$ and $a_1'$ mesons to be saturated. As it has
been done in previous determinations of other hadronic and generalized
form factors we have taken as an educated guess the half width rule
for the axial-vector isovector masses. The produced spread is fairly
consistent with the current experimental and lattice spread of values.

Most remarkably the errors in the axial form factor determined by the
axial-vector dominance and using the half width rule, while quite
generous, do not generally produce larger uncertainties in the
neutrino-nucleus scattering than the experimental differential cross
sections reported by the MiniBoone collaboration. We have also
provided evidence that the region of the axial form factor having most
impact in the MiniBooNE data is in the range $0.2 \lesssim Q^2
\lesssim 0.6 \,{\rm GeV}^2$, whereas fluctuations outside this regime tend
to be marginal. We stress that these features cannot be captured by
the conventional dipole parameterization.

Of course the minimal hadronic ansatz could be improved by adding
other poles from the PDG axial mesons compilation. In the case of
three poles, unlike the present case additional unknown information
such as e.g.  the coupling of the third meson to the nucleon is
needed. One could expect that the future neutrino data might be
accurate enough to pin down this extra parameter.

Our analysis has been carried out using the RFG model which for the
quasielastic region does not seem to miss effects which are larger
than the present uncertainty in the axial form factor using the AVMD
ansatz supplemented with the half-width rule. The role of additional
nuclear effects improving the present model will be presented in
further work.  The role played at higher energies by the equivalent
AVMD form factors and further nuclear mechanisms remains to be seen.

%An important ingredient of the present approach is the
%double binning of the thery mimicking the experimental measurements
%both in the angle and the neutrino energy. Given the predictive power
%of the present framework we ...

\section*{Acknowledgments}

We thank Pere Masjuan for collaboration in the very early stages of
this work. This work is supported by the Spanish Direccion General de
Investigacion Cientifica y Tecnica and FEDER funds 
(grant No. FIS2014-59386-P) and the
Agencia de Innovacion y Desarrollo de Andaluc{\'{\i}a} (grant
No. FQM225).

%\bibliographystyle{elsarticle-num}
%\bibliographystyle{unsrt}
%\bibliography{PLB.bib}
%\end{document}

%%%%%%%%%%%%%%%%%%%%%%%%%%%%%%%%%%%%%%%%%%%%%%%%%%%%%%%%%%%%%%%%%%%

\end{document}